\documentclass[12pt]{article}
\usepackage[tmargin=1in,bmargin=1in,lmargin=1.25in,rmargin=1.25in]{geometry}

\usepackage{setspace}
\usepackage{csquotes}
\usepackage{amsmath}
\usepackage{amsfonts}
\usepackage{stmaryrd}
\usepackage{dsfont}
\usepackage{graphicx}
\usepackage[font=small,labelfont=bf]{caption}
\usepackage{lipsum}
\usepackage{mwe}
\graphicspath{ {images/} }

\begin{document}

	\title{Do Simple Infinitesimal Parts Solve Zeno's Paradox of Measure?\footnote{Special thanks to Jeffrey Russell and Phillip Bricker for their extensive feedback on early drafts of the paper. Many thanks to the referees of \textit{Synthese} for their very helpful comments. I'd also like to thank the participants of Umass dissertation seminar for their useful feedback on my first draft.}}
	\author{Lu Chen}
	\date{ (Forthcoming in \textit{Synthese})}
	\maketitle
	
	\section*{Abstract}
	
	In this paper, I develop an original view of the structure of space---called \textit{infinitesimal atomism}---as a reply to Zeno's paradox of measure. According to this view, space is composed of ultimate parts with infinitesimal size, where infinitesimals are understood within the framework of Robinson's (1966) nonstandard analysis. Notably, this view satisfies a version of additivity: for every region that has a size, its size is the sum of the sizes of its disjoint parts. In particular, the size of a finite region is the sum of the sizes of its infinitesimal parts. Although this view is a coherent approach to Zeno's paradox and is preferable to Skyrms's (1983) infinitesimal approach, it faces both the main problem for the standard view (the problem of unmeasurable regions) and the main problem for finite atomism (Weyl's tile argument), leaving it with no clear advantage over these familiar alternatives. 
	\vspace*{4mm}
	
	\noindent \textbf{Keywords:} continuum; Zeno's paradox of measure; infinitesimals; unmeasurable regions; Weyl's tile argument
	
	\section{Zeno's Paradox of Measure}
	
    A continuum, such as the region of space you occupy, is commonly taken to be indefinitely divisible. But this view runs into Zeno's famous paradox of measure. If a finite line segment is indefinitely divisible, then it can be divided into infinitely many parts with the same length. In this case, if every part has a finite length, then the whole line segment would have an infinite length. But if every part has zero length, then the whole would have zero length. Either way, the whole would not have a finite length. This contradicts the assumption. (Skyrms 1983; see also Furley 1967)  

At its core, Zeno's paradox reveals the tension between two intuitive claims on the composition of a continuum: on the one hand, it is intuitive that every extended part of a continuum is further divisible, and given that, it is natural to consider an extensionless point to be an ultimate component of a continuum; on the other hand, an extended continuum cannot be exhaustively composed of extensionless points because zero sizes add up to zero. Leibniz, for example, was deeply puzzled by this tension and eventually concluded that continua are not real (Russell 1958).

We can break Zeno's paradox down into the following assumptions:

\begin{quote}
	\textsc{Infinite Divisibility}. A continuum can be divided into smaller and smaller parts without limit. 
	
	\textsc{Infinity Conditional}. If \textsc{Infinite Divisibility}, then a continuum can be divided into infinitely many parts of equal size.
	
	\textsc{Dichotomy.} Any part of a continuum has either zero size or at least a finite size.
	
	\textsc{Additivity.} The size of the whole is the sum of the sizes of its disjoint parts.
	
	\textsc{Zeros-Sum-To-Zero}. Zeros, however many, always sum up to zero.
	
\end{quote}

\noindent Since all these assumptions are required for the paradox to arise, each assumption provides a possible way to escape from the paradox. If we deny \textsc{Infinite Divisibility}, we will arrive at the view called \textit{finite atomism}, according to which the ultimate parts of a continuum are of finite sizes. If we deny \textsc{Infinity Conditional}, then we will come to \textit{the gunky view}, according to which a continuum does not have ultimate parts. This option can be traced back to Aristotle, who famously rejected \textsc{Infinity Conditional} by distinguishing between potential infinity and actual infinity. Finally, according to \textit{the standard view} in modern mathematics, a continuum is composed of uncountably many extensionless points, and therefore \textsc{Additivity} is false. Unfortunately, these approaches have their own problems (Section 2).  

Yet another possible approach to Zeno's paradox, following Skrym (1983), is to deny \textsc{Dichotomy} by claiming instead that a part of a continuum can have an infinitesimal size, a size that is non-zero but smaller than any finite size. In the past, many people have found the idea of an infinitesimal size unintelligible.\footnote{Infinitesimals first appeared in Democritus's work (450 B.C.E) only to be banished by Eudoxus (350 B.C.E). They reappeared during the invention of modern calculus, but were attacked and eventually abandoned in mainstream nineteenth-century mathematics. (Bell 2013)} But the situation started to change with the arrival of various systems of infinitesimals, such as Robinson's (1966) \textit{nonstandard analysis} and \textit{smooth infinitesimal analysis} developed by Lawvere (1980) and others. In this paper, I will explore the option of rejecting \textsc{Dichotomy} by extracting a new theory of continua, called \textit{infinitesimal atomism}, from the framework of nonstandard analysis.\footnote{I explore other technical approaches to infinitesimals such as smooth infinitesimal analysis elsewhere.} According to this theory, the ultimate parts of a continuum have an infinitesimal size. I will compare this approach with the standard view and finite atomism, the two more familiar approaches to Zeno's paradox that admit indivisible parts. My conclusion is that infinitesimal atomism, although having some attractive features, has no clear advantages over these familiar alternatives.

Note that infinitesimal atomism is not the only way to reject \textsc{Dichotomy}. Another possible approach is to have an infinitesimal version of the gunky view, according to which every part of a continuum is further divisible and some parts have infinitesimal sizes. I argue in Chen (manuscript a) that this new gunky approach has distinct advantages over current gunky theories. This essay is part of the project of examining what resources nonstandard analysis can provide for rethinking traditional paradoxes.

\section{Solutions and Their Problems}

The standard solution to Zeno's paradox, classically defended by  Gr\"unbaum (1973, 158-76), is based on standard analytic geometry. According to this view, a geometric line can be algebraically represented by the set of real numbers, with each real number representing a point on the line. This means that a line is exhaustively composed of points. The length of a point under standard measure theory (the Lebesgue measure) is zero, while the length of any line segment is a finite number. This is what we call \textit{the standard view}, according to which a line is exhaustively composed of points of zero length. 

The standard view violates \textsc{Additivity}, the principle that the size of the whole is the sum of the sizes of its disjoint parts. In standard analysis, there is a general definition of an arithmetic sum that satisfies \textsc{Zeros-Sum-To-Zero}, even in the uncountable case.\footnote{In standard analysis, even though we typically only sum up countably many numbers, summing up zeros is a notable exception. The following general definition of a sum over non-negative numbers indexed by an arbitrary set implies that zeros always sum up to zero. Let HUGE be an indexing set for some non-negative numbers. Let PAR be the set of partial sums of numbers indexed by finite subsets of HUGE. We define the sum of the numbers indexed by HUGE to be the least upper bound of PAR. It follows from this definition that zeros, even if  uncountably many, sum up to zero. This definition, when applies to a countable set of positive numbers, coincides with the more familiar notion of sum in standard analysis defined as the limit of partial sums of an infinite sequence.} However, according to the standard view, a point has zero length, and yet a line segment composed of uncountably many points has a finite length. Therefore, the length of the line segment is not the sum of the lengths of its constituent points.  

More straightforwardly, the standard view violates the principle that an extended whole cannot be composed of unextended parts. This principle---call it \textsc{Regularity}---is highly intuitive and remains an important consideration for evaluating a theory of continua.  Thus, the violation of this principle or \textsc{Additivity} is a substantial cost for the standard view. 

Apart from contradicting these intuitive principles, the standard view also leads to other measure-theoretical paradoxes, which are similar in spirit to Zeno's paradox. For instance, the Banach-Tarski paradox says that, assuming the axiom of choice, we can divide a sphere into finitely many disjoint parts, move them around rigidly (no stretching or squishing), and rearrange them to form \textit{two} spheres, each of which has exactly the same size as the original one (Wagen 1985; see also Forrest 2004).\footnote{There is also a more elementary analogous result in the one-dimensional case, which involves the construction of ``Vitali sets'' (see Skryms 1983, 238-9).} The measure theory of the standard view (the Lebesgue measure) satisfies finite additivity (and indeed countable additivity): for any finitely many disjoint parts of a continuum, the measure of their fusion is the sum of the measure of those parts. Moreover, the measure of a part does not change under rigid transformation. Therefore, standard measure theory cannot assign any measure to those parts: if they have any measures, it would be impossible for them to compose something twice as large as before without undergoing a change in measure. Call this problem ``the problem of unmeasurable regions." I will come back to this issue in Section 5.

The standard view not only faces \textit{a priori} objections but is also questionable on empirical grounds (for example, see Geroch 1972,  Arntzenius 2003). Moreover, physicists are actively designing empirical experiments to determine the structure of space (Hogan 2012).  What we philosophers can contribute, then, is to steer toward a clear view of what space could \textit{possibly} be like---stocking our warehouse with a rich store of possibilities, in the hopes that actuality might be one of them. 

Finite atomism is one of the main alternatives to the standard view.  This view, which says that a continuum is composed of finitely extended indivisible parts, was once considered unappealing because it violates \textsc{Infinite Divisibility}, the highly intuitive principle that every extended part can be further divided. But this view has become more popular because of favorable considerations from physics. It is even considered ``received wisdom" among physicists that a certain atomistic structure is required for reconciling quantum theory and general relativity (Maudlin 2015, 46). However, this view faces a troublesome argument given by Weyl (1949)---called \textit{Weyl's tile argument}---which purports to show that there are no natural distance functions over atomistic space that approximate Euclidean geometry. Since our space is approximately Euclidean at certain scales, Weyl's conclusion implies that finite atomism cannot describe our actual space. This problem, if unsolved, would leave finite atomism quite unattractive, for considerations from actual physics play an important role in motivating this view. I will explain Weyl's argument more in Section 6.

The gunky view, the view that every part of a continuum is further divisible, is another main alternative to the standard view. This approach is attractive because it does not violate highly intuitive principles such as \textsc{Infinite Divisibility} and \textsc{Regularity}. However, a natural development of this view turns out to be inconsistent (Arntzenius 2008; Russell 2008). Alternative gunky approaches are further discussed in Chen (manuscript a). 

Because of these problems, the situation is far from settled, and philosophers continue to look for alternative theories of continua or solutions to Zeno's paradox. The option of solving the paradox by appealing to infinitesimals has been contemplated from time to time but never developed into a full view of continua (for example, see Skyrms 1983). Without a clear and coherent view, it's hard to compare this option with other approaches to the paradox. So in this paper, I will first flesh out this option into a more developed view of continua---infinitesimal atomism---within the framework of nonstandard analysis. This view, I believe, is the most attractive atomistic theory of continua that we can extract from the mathematical ideas of nonstandard analysis. Notably, according to infinitesimal atomism, we can indeed sum up the lengths of infinitely many disjoint infinitesimal parts and get a finite length, although this kind of ``sum" is a new technical notion introduced by nonstandard analysis. 

Once we have a clear and coherent view in focus, we can start to evaluate it philosophically. Unfortunately, it turns out that the news for infinitesimal atomism is mostly bad. On the good side, the view indeed satisfies \textsc{Regularity} and other intuitive assumptions of Zeno's paradox to a certain extent (which I discuss in Section 4). On the bad side, it suffers \textit{both} from the main problem for the standard view (the problem of unmeasurable regions, which I discuss in Section 5)  \textit{and} from the main problem for finite atomism (Weyl's tile argument, which I discuss in Section 6). Neither of these difficulties is decisive, but they don't leave infinitesimal atomism with any clear advantages over more familiar alternatives.

	\section{Infinitesimal Atomism}
	
	In this section, I will develop the view \textit{infinitesimal atomism}, according to which a continuum is composed of ultimate parts of infinitesimal size, which I call ``minims,"  where infinitesimals are understood in the framework of nonstandard analysis. 

	In nonstandard analysis (NSA), the real line $\mathbb{R}$ is extended to the \textit{hyperreal line} $*\mathbb{R}$, which includes infinitesimals and infinite numbers along with the familiar real numbers.\footnote{My introduction of the hyperreal system is based on Goldblatt (1998). Note that there are multiple non-isomorphic hyperreal systems. The hyperreal system I introduce here is the smallest one. The surreal number system introduced by Conway (1976) is considered to be the largest hyperreal system.} A number is \textit{infinitesimal} iff its absolute value is smaller than any positive real number. A number is \textit{infinite} iff its absolute value is larger than any positive real number. On the hyperreal line, each real number is surrounded by a ``cloud" of hyperreal numbers that are infinitesimally close to it, called its \textit{monad}. The monads of different real numbers do not overlap. Moreover, we can do arithmetic with the hyperreal numbers just like with the standard real numbers. For example, for any positive infinitesimal $\delta$  and any infinite number $N$, we have $\sqrt{\delta}$, $\frac{1}{\delta}$, $\delta^2$, $\delta+N$, $\frac{\delta}{N}$, etc. More generally, the hyperreal numbers satisfy all the first-order truths about the real numbers in standard analysis.

The basic picture of infinitesimal atomism in the one-dimensional case is that the minims are lined up one by one with adjacent ones connected. Let $\Delta$ be the infinitesimal length of a minim. Then, it is natural to represent minims by consecutive intervals of length $\Delta$ on the hyperreal line, such as $...[0,\Delta),[\Delta, 2\Delta),...$\footnote{The half-open intervals, while they make a helpful intuitive picture, should not be taken too literally. I am using certain mathematical objects to represent parts of space, but not every mathematical feature corresponds to a genuine spatial feature. In particular, the asymmetric \textit{half-openness} does not correspond to any real feature of the minims that the intervals represent. Rather, the representation in question is a one-to-one correspondence between the half-open intervals and the minims that preserves length, order, and connectedness. Note that in the case of atomistic space, I take ``connectedness"  (or ``adjacency") to be primitive (see Forrest 1995 and Roeper 1997). Intuitively, two regions are connected if there is no gap between them.} For convenience, these intervals will be named by their left endpoints $...0,\Delta...$ Then, all minims are represented by members of the set $M=\{k\cdot\Delta: k\in *\mathbb{Z}\}$, where $*\mathbb{Z}$ is the hyperreal extension of the set of integers $\mathbb{Z}$ that satisfies all the first-order truths about $\mathbb{Z}$.

{\centering
\includegraphics[scale=0.7]{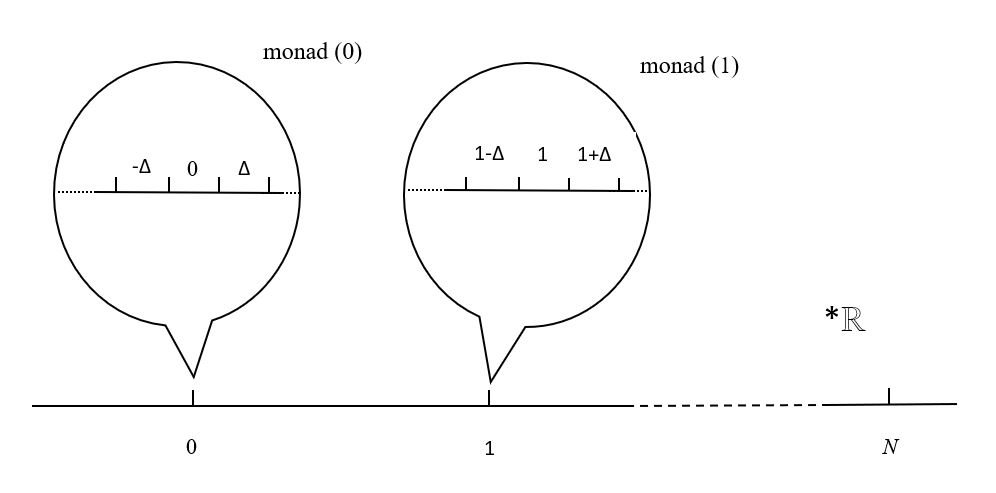}
\captionof{figure}{}}

 \noindent Note that although minims can be lined up one by one as shown in the figure,  there are actually \textit{uncountably} many minims between, for example, 0 and 1. In fact, for \textit{every} real number, there are uncountably many minims in its monad.\footnote{This can be derived from the theorem that every infinite hyperfinite set is uncountable (Goldblatt 1998, 141). In every monad, there are more than hyperfinitely many minims.}
	
	Minims are the smallest parts of the continuum, but what about others parts of the continuum? For brevity, let ``space" and ``region" be synonymous with ``continuum" and ``part of a continuum" respectively. Since space is entirely composed of minims, every region is also a fusion of minims. But I will not assume standard mereology and in particular, the principle of unrestricted composition:
	
	\begin{quote}
		\textsc{Unrestricted Composition.} Every collection of regions has a fusion. 
	\end{quote}
	
	\noindent The question of exactly what collections of minims should compose regions is a bit delicate, and I will postpone the discussion until Section 5. But for now, we can at least assume that collections of regions of a special kind---\textit{hyperfinite} collections of regions---have fusions.
	
	The notion of ``hyperfiniteness" is a distinct notion of cardinality in NSA. When extending $\mathbb{R}$ to $*\mathbb{R}$, the set of all natural numbers is extended to the set of \textit{hypernatural} numbers, which preserves all the first-order truths about natural numbers in standard analysis. Just as every real number is bounded by a natural number, every hyperreal number is bounded by a hypernatural number. Since there are infinite hyperreal numbers, there are also infinite hypernatural numbers. Let $N$ be an infinite hypernatural number. Consider the set $\{1,2,...,N\}$. How many elements does it have? The answer is $N$---according to the new notion of \textit{hyperfinite cardinality} in NSA. Just as in standard analysis the set $\{1,2,...,n\} (n\in\mathbb{N})$ has a cardinality of $n$, in NSA the set $\{1,2,...,N\}$ has a hyperfinite cardinality of $N$. Note that the notion of hyperfinite cardinality is finer-grained than the notion of cardinality in standard set theory. According to standard set theory, if $N$ is infinite, then $\{1,2,...,N\}$ and $\{1,2,...,N+1\}$ have the same cardinality. But in NSA, they do not have the same hyperfinite cardinality. Two sets have the same hyperfinite cardinality just when there is a bijective \textit{internal} function between them, where ``internal" means being expressible in the language of standard analysis.\footnote{Note that the language of standard analysis includes quantifiers over \textit{sets} and \textit{functions}, as in this definition of hyperfiniteness (Goldblatt 1998, 178-80). These quantifiers also receive non-standard ``internal" interpretations in the hyperreal system (Goldblatt 1998, 168-70).} (For ``hyperfiniteness," see Goldblatt 1998, 178-81; for ``internal entities," see 166-73.)
	
	Given the notion of hyperfinite cardinality, we assume the following composition principle:	
	
	\begin{quote}
		\textsc{Hyperfinite Composition.} Every hyperfinite collection of regions has a fusion. 
	\end{quote}
	
	\noindent  Let's assume that $\Delta$ is a multiplicative inverse of an infinite hypernatural number. Then, according to \textsc{Hyperfinite Composition}, the minims between $0$ and $1$, which are represented by $0, \Delta,...,\frac{1}{\Delta}\cdot \Delta$, compose a region, because $\{0, \Delta,...,\frac{1}{\Delta}\cdot \Delta\}$ is hyperfinite. But this principle does not tell us whether, for instance, the minims in the monad of $0$ compose a region, because the set of those minims is not hyperfinite---it's not hyperfinite because there is no maximal hypernatural $N$ such that $N\cdot \Delta$ is infinitesimal.

The measure theory for infinitesimal atomism is especially simple: we can find the measure of a region just by counting the minims it contains. The trick that makes this work is that we have the notion of \textit{hyperfinite sum} in NSA. Just as every finite sequence of numbers in standard analysis has a sum, in NSA, every hyperfinite sequence of numbers has a hyperfinite sum. As a simple example, if we add up the infinitesimal length $1/N$ $N$-many times, then the result is one. Note that this notion is very different from the notion of ``infinite sum" in standard analysis (Footnote 3).

\begin{quote}
	\textsc{Hyperfinite Measure}. For any hyperfinite collection of minims, the measure of their fusion is equal to the hyperfinite sum of the measures of those minims.\footnote{For those who are mathematically informed, the measure is closely related to Loeb measure in NSA. The main difference is that Loeb measure is real-valued and is defined over a $\sigma$-algebra which includes the domain of the hyperreal measure as a proper subset.}
\end{quote}

\noindent Then, the fusion of $N$-many minims has the measure of $N\cdot \Delta$.
	
	One desirable feature of the measure is that it generally gives familiar results: it approximates the Lebesgue measure (the standard measure over the real line) up to infinitesimal differences. For example, consider again the set of minims between $0$ and $1$: $\{0, \Delta, ...,(\frac{1}{\Delta}-1)\cdot \Delta\}$, which has a hyperfinite cardinality of $\frac{1}{\Delta}$. Thus the fusion of those minims has a measure of $\frac{1}{\Delta}\cdot \Delta=1$, which is equal to the Lebesgue measure of the interval [0,1]. (For more general results, see Goldblatt 1998, 215-7.)

	Although all measurable regions  are fusions of hyperfinitely many minims, the measure theory is actually very rich. The \textit{shadow} of a hyperreal number (or its corresponding minim), if it exists, is the real number that is infinitesimally close to it (note that an infinite hyperreal does not have a shadow). For \textit{every} Lebesgue-measurable set $X$, we can find minims whose shadows are elements of $X$ such that the hyperreal measure of the fusion of those minims is infinitesimally close to the Lebesgue measure of $X$ (212-3, 216-7). Let each Lebesgue-measurable set be mapped to such a measurable region. Then sets with different Lebesgue measures are mapped to regions with different hyperreal measures. Meanwhile, it's clear that regions with different hyperreal measures do not correspond to sets with different Lebesgue measures in an analogous sense. For example, a region composed of a single minim and a region composed of two each correspond to a singleton of a point, which has Lebesgue measure zero. This means that we have an injective but not surjective map from Lebesgue-measurable sets to measurable regions that preserves their measures up to infinitesimal differences. In this sense, infinitesimal atomism has an even richer measure than the standard measure! 
	
	\section{Revisiting Zeno's Paradox}

	Infinitesimal atomism avoids Zeno's paradox by dispensing with \textsc{Dichotomy}, the principle that implies there are no infinitesimal regions. But whether the view solves Zeno's paradox satisfactorily depends on how well it satisfies other intuitive assumptions in the paradox. In this section, I will give a preliminary examination of this question and leave some of the more complicated issues to Section 5.

	First, consider \textsc{Infinite Divisibility}. In the context of hyperreals, we can distinguish between the following two versions:
	
	\begin{quote}
		\textsc{Finite Divisibility}. Every finitely extended region can be divided into smaller parts.
		
		\textsc{General Divisibility}. Every extended region can be divided into smaller parts.  
	\end{quote} 
	
	\noindent The standard view satisfies both principles. Finite atomism satisfies neither. Infinitesimal atomism satisfies \textsc{Finite Divisibility} but not \textsc{General Divisibility}, since according to the view, any finitely extended region can be divided into infinitely many minims, but each minim is an indivisible extended region.  
	
	One might think that regardless of whether \textsc{Finite Divisibility} holds, it is equally bad to deny \textsc{General Divisibility}. The intuition behind this principle can be spelled out through the following argument given by Zimmerman (1996). Every extended region has a size and a shape. Every region that has a size and a shape has proper parts---for instance, a unit square must have a left half and a right half, each of which is a half-unit rectangle. Thus, every extended region must have proper parts and hence be divisible.\footnote{There is a related worry of arbitrariness that I do not discuss in this paper. The worry is that it is arbitrary that a minim has a length of a particular infinitesimal. As pointed out by the anonymous referee, Reeder (2015) discussed this worry and suggested that we should represent a minim by a monad instead. I prefer the approach in this paper because it allows for a richer measure theory with attractive features. We can add up the lengths of minims to obtain an infinitesimal or a finite length, and the measure of any measurable region is the sum of its ultimate parts. In contrast, monads are unmeasurable, and we cannot add up their lengths. In fact, the measure theory of Reeder's approach would be less rich than the standard measure: they are otherwise the same except that points of length zero are replaced by unmeasurable monads.} 
	
	Let's grant that it is intuitive that every extended region has a proper part. One thing we could say in reply is that this intuition is reliable only when the region in question is finite. When it comes to the infinitesimals, our intuition need not be reliable: we have a technical theory about them, but we don't necessarily have an intuitive grasp of them. After all, infinite numbers are notoriously counterintuitive: Hilbert's hotel is fully occupied and yet can accommodate infinitely more! So it might well be the case that our intuition is equally unreliable in the case of infinitesimals, which are just reciprocals of infinite numbers. In this sense, the violation of \textsc{General Divisibility} in infinitesimal atomism might be more acceptable than the violation of \textsc{Finite Divisibility} in finite atomism. 
	
	The case of \textsc{Additivity} is tricky. On the one hand, the measure theory satisfies the following version of additivity:\footnote{Note that this is a distinct feature of infinitesimal atomism that is not present in, for example, Skryms's (1983) suggestion. The measure theory Skyrms suggested is only \textit{finitely additive} (241-2).} 
	
	\begin{quote}
		
		\textsc{Hyperfinite Additivity}. For hyperfinitely many measurable disjoint regions, the measure of their fusion is the (hyperfinite) sum of the measures of those regions.
	\end{quote}
	
	\noindent This is true because the fusion of any hyperfinitely many measurable disjoint regions is a fusion of hyperfinitely many minims.\footnote{This is because the union of hyperfinitely many disjoint hyperfinite sets is still hyperfinite, just like the union of finitely many disjoint finite sets is still finite.} The hyperfinite number we obtain by counting the minims in the fusion is the same number as we get by counting the minims in each region and then adding them together. 
	
	On the other hand, the original principle of \textsc{Additivity} is not restricted to hyperfinite collections of regions. Since a non-hyperfinite collection of regions may not have a measurable fusion at all, ``the measure of the fusion" in \textsc{Additivity} can be an empty description. How should we understand this principle then? Here's a natural option:
	
	\begin{quote}
		\textsc{Weak Additivity.} For arbitrarily many measurable disjoint regions, if they have a fusion, and if their fusion is measurable, then the measure of their fusion is the sum of the measures of those regions.

	\end{quote}
	
	\noindent Infinitesimal atomism satisfies this principle.\footnote{In contrast, Skyrms's (1983) approach violates \textsc{Weak Additivity}. The fact that Skyrms's suggested measure is only finitely additive (see Footnote 11) means that for any region of a finite size, its measure is \textit{not} the sum of the measures of its composing points, since every finite region contains uncountably many points. This feature, I think, makes his approach less attractive than mine. The violation of the intuitive principle of \textsc{Weak Additivity} alone is a cost. Moreover, this violation also undermines the motivation for \textsc{Regularity}. If the measure of an extended region is not determined by the measures of its ultimate parts, does it matter whether the ultimate parts have measure zero? It doesn't seem so bad to have an unextended point \textit{per se}. What's bad rather seems to be the failure of \textsc{Additivity} when those unextended points compose an extended region. So, it is not clear that Skyrms's approach is an improvement over the standard view.} For arbitrarily many measurable disjoint regions, either they are hyperfinitely many, or their fusion is not measurable, since every measurable region is a fusion of hyperfinitely many measurable regions. If the regions are hyperfinitely many, the principle is reduced to \textsc{Hyperfinite Additivity}. If their fusion is unmeasurable, then the principle is vacuously true. Either way, the principle is satisfied. In contrast, the standard view does not satisfy this principle, for according to the standard view, any finite line segment is a fusion of uncountably many points of zero size, but zeros sum up to zero.

	Having \textsc{Weak Additivity} is desirable, but not fully satisfactory. A stronger version of additivity would require that the fusion of arbitrarily many measurable disjoint regions is also measurable:
	
	\begin{quote}
		\textsc{Strong Additivity.} For arbitrarily many measurable disjoint regions, if they have a fusion, then their fusion is measurable, and the measure of their fusion is the sum of the measures of those regions.\footnote{This version of additivity is as strong as it can get without building in \textsc{Unrestricted Composition}, the principle that any regions have a fusion. Under \textsc{Unrestricted Composition}, \textsc{Strong Additivity} entails the full-blown version of additivity: for arbitrarily many measurable disjoint regions, they have a fusion, their fusion is measurable, and the measure of their fusion is the sum of the measures of those regions.}
	\end{quote}
	
	\noindent However, whether infinitesimal atomism satisfies \textsc{Strong Additivity} depends on what composition principles hold. I will take up this question in the next section.

So far, infinitesimal atomism does reasonably well on meeting the intuitive assumptions of Zeno's paradox. It satisfies \textsc{Regularity}, since every region is extended. It also satisfies \textsc{Infinite Divisibility} in the finite case, though not in general. Moreover, the view satisfies \textsc{Weak Additivity} and \textsc{Zeros-Sum-To-Zero}. Whether it satisfies \textsc{Strong Additivity}, or at what cost, will be discussed in the next section---where the bad news begins.

	\section{The Problem of Non-Measurable Regions}

	Suppose \textsc{Unrestricted Composition} is true: every collection of minims has a fusion. Then some regions are not measurable.\footnote{Here I assume that the fusion of regions is also a region. This follows from my stipulation that ``region" simply means a part of the continuum, together with the standard mereological principle that the fusion of parts of X is still a part of X.} This could happen, for example, when the region is the fusion of countably infinitely many disjoint regions, each of which is measurable. Imagine that you are walking along a straight line from the minim 0. You first walked 1/2 mile, then 1/4 mile, then 1/8 mile, and so on.\footnote{More precisely, the sequence of distances you walked is $\langle 1/2^n\rangle$, where $n$ ranges over all \textit{natural} numbers.} How many miles have you walked in total? The answer is ``Unmeasurable." This is because the minims that are infinitesimally close to the minim 1 are not included in your journey. The fusion of these minims does not have a measure because these minims are not hyperfinitely many. So the total distance you have traveled is not measurable. Also, a region composed of minims in a monad is not measurable, since those minims are not hyperfinitely many (Section 3). 
	
	Indeed, if we assume \textsc{Unrestricted Composition}, then there are many more unmeasurable regions than those under the standard view. For every set of real numbers, let's map it to the fusion of all minims that are infinitesimally close to those real numbers.  All such fusions are unmeasurable, while sets of real numbers are generally Lebesgue measurable.\footnote{Like a monad, the sets of the minims in question are not \textit{internal sets}, sets that are characterizable in the language of standard analysis.} For example, the singleton set of a real number, which has a Lebesgue measure of zero, is mapped to the fusion of the minims in its monad, which is unmeasurable. The interval [0,1] is mapped to the fusion of the minims in the hyperreal interval $*[0,1]$ together with 0's and 1's monads, which is again unmeasurable.  Thus we have an injective but not surjective map from non-Lebesgue-measurable sets of real numbers to unmeasurable regions that preserves mereological relations. It is unclear---and unlikely---that there is such a map from unmeasurable regions to non-Lebesgue-measurable sets. After all, even within a single monad, there are already uncountably many unmeasurable regions!\footnote{Every region composed of countably infinitely many minims is unmeasurable, and every monad contains uncountably many countably infinite collections of minims.}

	Therefore, if every collection of minims has a fusion, then like the standard view, infinitesimal atomism faces the problem of unmeasurable regions. The situation seems even worse for infinitesimal atomism, since there are more unmeasurable regions. Moreover, one strategy to get rid of non-Lebesgue-measurable sets in standard analysis is to reject the axiom of choice, the principle that for any collection of nonempty sets, there is a set that has exactly one element from each of those sets (Solovay 1970). But this option is not available for infinitesimal atomism: the existence of unmeasurable regions does not require the axiom of choice.\footnote{The ultrafilter construction of the hyperreal system requires only \textit{the Boolean prime ideal theorem}, which is strictly weaker than the axiom of choice. Thanks to an anonymous referee for pointing out this difference.}
	
	For the sake of argument, let's assume that it is problematic to have unmeasurable regions. Does infinitesimal atomism have a better response than the standard view? In particular, can we deny \textsc{Unrestricted Composition} instead? Although those who are convinced of this principle (such as Lewis 1991 and Bricker 2015) are unlikely to be moved by considerations about measure theory alone, this option would still be attractive if something principled and intuitively relevant can be said about what collections of regions have fusions. The following principle is a natural candidate if we want to avoid unmeasurable regions: 
	
	\begin{quote}
		\textsc{Hyperfinite Composition$\dagger$.} A collection of regions has a fusion if and only if it's hyperfinite.\footnote{To adopt this principle, we must impose a further constraint on the set of minims $M$: it must be bounded by hyperreal numbers. Otherwise the minims in $M$ would not be hyperfinitely many, and thus would not compose a region.}
	\end{quote}
	
	\noindent If we adopt \textsc{Hyperfinite Composition}$\dagger$, it will follow that all regions are measurable.  But is the condition of hyperfiniteness intuitively relevant for composition? Here's one line of reasoning that may encourage such an idea. In fact, the hyperfinite subsets of $M$ are precisely those \textit{internal} subsets of $M$, namely those that are expressible in the language of standard analysis.\footnote{Here, we assume that $M$ is bounded by hyperreal numbers (Footnote 20). Also see Goldblatt (1998, 166-73) for more information about internal sets.}  Let $\phi$ be a schematic variable for any one-place formula in a suitable language, and let Fus$_\phi z$ mean that $z$ is a fusion of all the things that satisfy $\phi$.\footnote{Fus$_\phi z$ can be defined in terms of parthood in a first-order language (let ``$Pxy$" mean $x$ is a part of $y$): Fus$_\phi z$ iff $\forall x(\phi x\to Pxz)\wedge \forall y(\forall x(\phi x\to Pxy)\to Pzy)$. (Cotnoir and Varzi 2018)}  Consider the following schematic way of capturing \textsc{Unrestricted Composition} (Cotnoir and Varzi 2018):
	
	\begin{quote}
		($\Phi$)  $\exists x \phi(x) \to \exists z$Fus$_\phi z$.
	\end{quote}
	
	\noindent If we take $\phi$ as only ranging over the formulas in the language of standard analysis (call it $\mathcal{L}$), then given that all and only hyperfinite subsets of $M$ are $\mathcal{L}$-expressible, $\Phi$ is equivalent to \textsc{Hyperfinite Composition}$\dagger$.
	
	 But why should we restrict $\phi$ in $\Phi$ to the $\mathcal{L}$-formulas? Such a restriction is reasonable only if we think the nonstandard model only has internal entities. For example, one might think that, because the countably infinite set of minims $\{\Delta, 2\Delta,...\}$ is not $\mathcal{L}$-expressible, there is no such set in the nonstandard model. Therefore, the minims $\Delta, 2\Delta,...$ do not compose a region. But such a connection between existence and $\mathcal{L}$-expressibility is indefensible. After all, the notion of \textit{infinitesimal} is not $\mathcal{L}$-expressible.\footnote{Recall that infinitesimals are those whose absolute values are smaller than any positive real number. But in NSA, the set of positive hyperreal numbers satisfies the exact same first-order truths as the set of positive real numbers in the language of standard analysis. So the set of positive real numbers is not expressible, which means that the set of infinitesimals is not expressible either.} But we still think the notion of infinitesimal intelligible and that there is, for example, a collection of regions infinitesimally close to zero---at least if we want to take infinitesimal atomism seriously. So I don't think that a restriction of composition to \textit{internal} collections of regions is well-motivated.

	Therefore, infinitesimal atomism doesn't appear to have better options than the standard view in answering the problem of unmeasurable regions. Like the standard view, infinitesimal atomism leads to the result that there are unmeasurable regions,  and therefore also violates \textsc{Strong Additivity}. Note that I am \textit{not} arguing it is necessarily bad to have unmeasurable regions---the point is just that the problem of unmeasurable regions cannot be the reason for favoring infinitesimal atomism over the standard view.
	
	Notice that \textit{finite} atomism does not face this problem. For every region, it is either composed of finitely many atoms or else of infinitely many. If a region is composed of finitely many atoms, which have finite sizes, the region has a finite size; otherwise it has a size of positive infinity $+\infty$. So every region is measurable. 
	
	\section{Weyl's Tile Argument}

	Weyl's tile argument purports to show that if finite atomism is true, then the Pythago-rean theorem is not even approximately true. So, as long as we hold onto the Pythagorean theorem as an approximate law of geometry, finite atomism is false. Philosophers and physicists have found the argument challenging. In what follows, I will first explain Weyl's tile argument and its underlying assumptions and only turn to infinitesimal atomism at the end. Weyl wrote:

	\begin{quote}
		How should one understand the metric relations in space on the basis of this idea? If a square is built up of miniature tiles, then there are as many tiles along the diagonal as there are along the side; thus the diagonal should be equal in length to the side. (Weyl 1949, 42)
	\end{quote}

	\noindent For example, consider the region composed of $4\times 4$ square tiles shown in Figure 2.\footnote{My presentation of the argument draws on Salmon (1980).}   $A,B,C$ are corner tiles. The side of the square $AC$ is four units long. There are four tiles along the diagonal $BC$, which, according to Weyl, implies that $BC$ is four units long. But if the Pythagorean theorem is approximately true, then $BC$ should be about 5.7 units long. Adding more tiles does not help. If the square is composed of $8\times 8$ tiles (Figure 3), the diagonal is still as long as the side---both are eight units long.

	\begin{figure}
	\begin{minipage}{0.45\textwidth}
		\centering
		\includegraphics[width=0.7\textwidth]{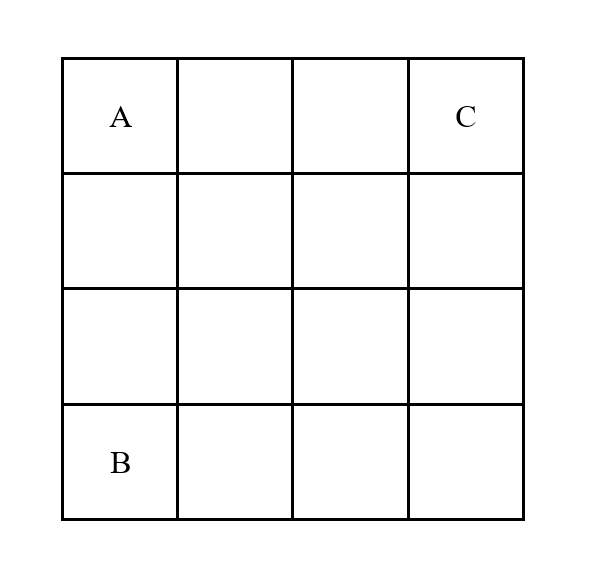} 
		\caption{}
	\end{minipage}\hfill
	\begin{minipage}{0.45\textwidth}
		\centering
		\includegraphics[width=0.75\textwidth]{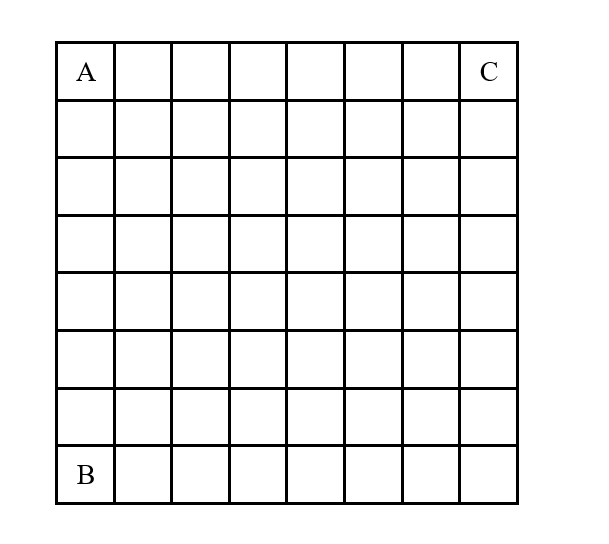} 
		\caption{}
	\end{minipage}
	
\end{figure}

	The argument relies on an important assumption about length in atomistic space: the length of a ``line segment" is equal to the number of tiles the ``line segment" contains. This assumption can also be put in terms of distance:\footnote{Although it is possible to distinguish between ``length," a \textit{measure} defined over regions of space, and ``distance," a \textit{metric} defined over pairs of points or atoms, it is standardly assumed that the argument can be put in terms of distances.} 
	
	\begin{quote}
		\textsc{Distance.}  The distance between any two tiles is equal to the number of tiles between them.
	\end{quote}
	
\noindent What is the rationale for \textsc{Distance}? As Bricker (1993) pointed out, our best theory of physics supports a \textit{path-dependent} \textit{local} account of distance: the distance between any two points is the length of the shortest path between them, and the length of a path is determined by the local metric properties of the points along the path. Although the standard definition of path does not directly apply to atomistic space, we can define ``path" in atomistic space as a fusion of minims in a ``chain of adjacency."\footnote{In standard differential geometry, a path between two points is a continuous function from the real interval [0,1] to the space, which takes 0 to one of the two points and 1 to the other one.} More precisely, a \textit{path} between minims $M_1, M_n$ is the fusion of minims $M_1,M_2,...,M_n$ such that  $M_k$ and $M_{k+1}$ are connected for every $k=1,...,n-1$.\footnote{\textit{Connectedness} is a primitive topological notion for finite atomism and infinitesimal atomism.} Given this definition, Weyl's tile argument can be understood as implicitly assuming that two tiles are connected just in case they are horizontally, vertically, or diagonally adjacent. It follows that the four tiles along the diagonal in Figure 1 compose a path.\footnote{Another intuitive option is to stipulate that two tiles are connected \textit{iff} they are horizontally or vertically adjacent. Under this option, the diagonal $BC$ is the zigzag region shown below. This option is similarly problematic.

{	\centering
	\hspace*{35mm}\includegraphics[width=0.3\textwidth]{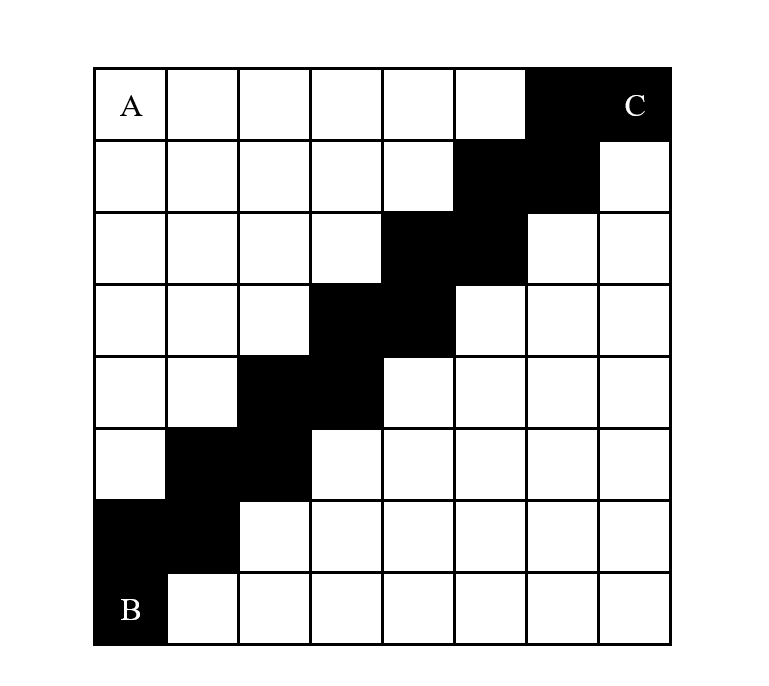}}

	} Furthermore, we can take Weyl's tile argument as assuming that, in atomistic space, the size of a minim (which is one unit) is the only primitive local metric property. Then, according to the path-dependent local account of distance, the length of a path in atomistic space is equal to the number of minims the path contains. Given that the distance between any two minims is equal to the length of the shortest path, \textsc{Distance} follows.
	
Suppose that finite atomism is subject to Weyl's tile argument. In that case, can infinitesimal atomism do better? The answer is no. Let $N$ be an infinite hypernatural number: Consider a region composed of $N\times N$ tiles. Let each tile  be represented by a pair of hypernatural numbers. For example, the tiles contained in a side can be represented by $(1,1), (1,2),...,(1,N)$. The tiles contained in a diagonal can be represented as $(1,1), (2,2),...,(N,N)$. Both sets of tiles clearly have the hyperfinite cardinality of $N$. Therefore, if we assume \textsc{Distance} applies to hyperfinite cardinality, the same departure from the Pythagorean theorem ensues.\footnote{Skyrms's continuum might be able to escape this argument. Although Skryms did not explicitly define a distance function over the continuum, he can resort to the standard distance function over standard space because Skyrms's continuum is isomorphic to standard space. The infinitesimal measure of a ``point" does not play any role in determining distances.} 
	
As before, I am not saying that Weyl's tile argument is devastating for infinitesimal atomism, but rather that it does not provide a reason to favor infinitesimal atomism over ordinary finite atomism. (In fact, I think finite atomism has a good response to Weyl's tile argument, which I defend in Chen (manuscript b); see also Forrest 1995, Van Bendegem 1987, 1997 for other responses.)

	\section*{Conclusion}
	
	In this paper, I have developed a theory of continua with a non-standard measure, infinitesimal atomism, in response to Zeno's paradox of measure. Among the principles Zeno's paradox relies on, \textsc{Dichotomy} is denied because the ultimate parts of continua have an infinitesimal measure, while other principles are satisfied to some degree. Most notably, the theory satisfies \textsc{Weak Additivity}, the principle that the measure of every measurable region is the sum of the measures of its disjoint parts. The measure theory is very rich and approximates the familiar Lebesgue measure over standard space. \textsc{Regularity} is satisfied, in which respect infinitesimal atomism is more intuitive than the standard view.
	
	However, it turned out that the theory does not score well against its main competitor, finite atomism, the familiar view that continua are composed of finitely extended atoms. Finite atomism has most of the benefits of infinitesimal atomism, such as \textsc{Regularity} and \textsc{Weak Additivity}. Furthermore, it satisfies \textsc{Strong Additivity} and is free of the problem of unmeasurable regions, which troubles both the standard view and infinitesimal atomism. Weyl's tile argument is the main drawback for finite atomism, but infinitesimal atomism has no special resources for answering it that are unavailable to finite atomism. Infinitesimal atomism does have two distinctive advantages over finite atomism: first, it satisfies \textsc{Finite Divisibility}, the principle that every finite region can be further divided into smaller regions, and second, it has a richer measure than either of its competitors. But all things considered, it's hard to say that these advantages count decisively in its favor.
	
	 Therefore, if we are not satisfied with the standard view and finite atomism, infinitesimal atomism provides no better consolation, and we must continue looking for alternatives.

\end{document}